\documentclass[fleqn,useAMS, usenatbib]{mnras}

\usepackage[dvips]{graphics}
\usepackage{graphicx}
\usepackage{amsmath}
\usepackage{epsfig}
\usepackage[latin1]{inputenc}

\usepackage[T1]{fontenc}
\usepackage{ae,aecompl}

\usepackage{verbatim}

\usepackage{multicol}

\def\lesssim{\la}

\newcommand\NS{N\!S}
 \newcommand{\be}{\begin{equation}}
 \newcommand{\ee}{\end{equation}}
 \newcommand{\ba}{\begin{eqnarray}}
 \newcommand{\ea}{\end{eqnarray}}
 \newcommand{\bs}{\begin{subequations}}
 \newcommand{\es}{\end{subequations}}
 \newcommand{\erbold}{{\bf{r}}}
 
 \newcommand{\eRbold}{{\bf{R}}}
 \newcommand{\Rbold}{{\bf{R}}}

\title[]{Numerical Simulation of Tidal Evolution of a Viscoelastic Body Modelled with a Mass-Spring Network}

\author[Frouard et al.]
{Julien Frouard$^{1}$\thanks{Contact e-mail: \href{mailto:jfrouard@federatedit.com}{jfrouard@federatedit.com}}, Alice C. Quillen$^{2}$, Michael Efroimsky$^1$, David Giannella$^2$ \\
$^1$US Naval Observatory, 3450 Massachusetts Ave NW, Washington DC 20392 USA \\
$^2$Department of Physics and Astronomy, University of Rochester, Rochester, NY 14627 USA \\
}

\begin{document}

\maketitle

 \begin{abstract}
 We  use a damped mass-spring model within an N-body code to simulate the tidal evolution of the spin and orbit of a self-gravitating viscoelastic spherical body moving around a point-mass perturber. The damped mass-spring model represents a Kelvin-Voigt viscoelastic solid. We measure the tidal quality function (the dynamical Love number $\,k_2\,$ divided by the tidal quality factor $\,Q\,$) from the numerically computed tidal drift of the semimajor axis of the binary. The shape of $\,k_2/Q\,$, as a function of the principal tidal frequency, reproduces the  kink shape predicted by Efroimsky (2012a; {\it{CeMDA}} 112$\,:\,$283) for the tidal response of near-spherical homogeneous viscoelastic rotators. We demonstrate that we can directly simulate the tidal evolution of spinning viscoelastic objects. 
In future, the mass-spring N-body model can be generalised to inhomogeneous and/or non-spherical bodies.

 \end{abstract}



\section{Motivation and Plan\label{sec1}}

 The analytical theory of tidal interaction of solid bodies has a long and rich history~---~from the early mathematical development by
 \citet{darwin},$\,$\footnote{~Darwin's work is presented, in the modern notation, in \citet{ferrazmello08}.} to its generalisation by \citet{kaula64}, to an avalanche of more recent results. Verification of tidal theories through direct measurements is not easy because the tidal evolution is slow and requires either  high observational precision \citep{wb2015} or an extended observational time span \citep{laineyetal2012}. Purely analytical theories of tidal evolution describe homogeneous or, at best, two-layered (as in \citealt{remus15}) near-spherical bodies of linear rheology. This makes numerical simulations attractive  (e.g., \citealt{henning14}) as  they   may make it possible to explore the tidal evolution of more complex objects. 

Here we explore  computations that treat the solid medium as a set of mutually gravitating massive particles connected with a network of damped massless springs. Because of their simplicity and speed (compared to more computationally intensive grid-based or finite element methods),  mass-spring computations are a popular method for simulating soft deformable bodies (e.g., \citealt{nealen06}). \citet{ostoja02} and \citet{kot14} have shown that mass-spring systems can accurately model elastic materials.  As we shall show, viscoelastic response and gravitational forces can be incorporated into the mass-spring particle-based simulation technique. We demonstrate  numerical simulations of the tidal response of a spinning homogeneous spherical body exhibiting spin-down or up and associated drift in semi-major axis, without using an analytical tidal evolution model. Then we compare the outcome of our numerics with analytical calculations. The results are similar, justifying  our numerical approach. Our numerical model may in future be used to take into account more complex effects that cannot be easily computed by analytical means (like inhomogeneity, compressibility, or a complex shape of the tidally perturbed body).

\section{Tides in a Kelvin-Voigt Viscoelastic Solid\label{sec2}}

\subsection{How tides work}

Consider an extended spherical body of mass $\,M\,$ and radius $\,R\,$, tidally deformed by a perturber of mass $\,M^*\,$ residing at a position $\,{\erbold}\,$, where $\,|\erbold|\geq R\,$. The binary's orbit has the semi-major axis $\,a\,$ and the mean motion $\,n = \sqrt{G(M + M^*)/a^3}\,$, where $\,G\,$ is the gravitational constant. For a distant perturber ($\,a\gg R\,$), the quadrupole part in the Fourier expansion (\ref{L5}) for the perturbing potential $\,W\,$ is dominant. This part comprises several terms. Of these, the term called $\,${\it{semidiurnal}}$\,$ is usually leading.$\,$\footnote{~See Section \ref{1.5} of the Appendix.} This term is a function of the principal tidal Fourier mode $\,\omega_{2200}\,$ which we denote simply as $\,\omega\;$:
 \ba
 \omega~\equiv~\omega_{2200}\,=~2(n~-~\dot \theta)\quad,
 \label{eqn:omega_simple}
 \ea
 where $\,\theta\,$ and $\,\dot\theta\,$ are the body's rotation angle and spin rate in the equatorial plane. (See the equation (\ref{L14}) in the Appendix.)

 The secular part of the semidiurnal term of the polar tidal torque acting on the body is
 \ba
 \langle {\cal{T}}^{(z)}_{2200}\rangle\;=\;\frac{3}{2}\;G\;M^{*2}\;\frac{R^5}{a^6}\;k_2(\omega)\;\sin\epsilon_2(\omega)\quad,
 \label{eqn:T_simple}
 \ea
 where $\,k_2(\omega)\,$ and $\,\epsilon_2(\omega)\,$ are the quadrupole dynamical Love number and the quadrupole phase lag, both taken at the semidiurnal frequency given by the above expression (\ref{eqn:omega_simple}). The product $\,k_2(\omega)\,\sin \epsilon_2(\omega)\,$ is often called $\,${\it{the quality function}}$\,$ \citep{makarov12, makarovMoon, efroimsky15} or, sometimes, $\,${\it{kvalitet}}$\,$ \citep{makarovSemiliquid,frouard}. 

 When the inclination and eccentricity are small, conservation of angular momentum gives an estimate of the secular drift rate of the semi-major axis (see the equation  (\ref{L17}) in the Appendix):
\begin{eqnarray}
  \frac{\dot a}{n\,a} &=&~-\frac{2\,{\cal{T}}^{(z)}_{2200}\,a}{G M^* M} \nonumber \\
 &=& - 3 \left( \frac{M^*}{M} \right)\left( \frac{R}{a} \right)^5\;k_2(\omega)\;\sin\epsilon_2(\omega) \quad.\quad
 \label{eqn:daa}
 \end{eqnarray}
 Below we compute the drift rate of the semi-major axis $\,\dot a\,$ through a direct numerical simulation, and then compare the result with that obtained analytically from the tidal theory using a homogenous Kelvin-Voigt viscoelastic rheology. This comparison will  demonstrate that we can directly simulate the tidal evolution of viscoelastic objects with  a mass-spring model.

 \subsection{The shape of the quality function}

 As is demonstrated in the Appendix, the Fourier decompositions of the disturbing potential $\,W\,$, the tidal response potential $\,U\,$, and the tidal torque $\,{\cal{T}}\,$ comprise terms that are numbered with the four indices $\,lmpq\,$. An $\,lmpq\,$ term of the torque is proportional to the quality function
 $\,k_l(\omega_{lmpq})\;\sin\epsilon_l(\omega_{lmpq})\,$. The quality function  depends on the degree {\emph{l}}, the composition of the body, the rheology of its layers, and the  size and mass of the body. The size and mass are important, because the tidal response is defined not only by the internal structure and rheology, but also by self-gravitation.

 The process of deriving the quality function for any linear viscoelastic rheology is described in \citet{efroimsky15}. Here we provide a short account of that derivation. We begin with the expression for the static Love number for an homogeneous, incompressible, self-gravitating elastic sphere,
 \begin{equation}
 k_l^{(static)}\,=\;\frac{3}{2(l-1)}\;\frac{1}{1+B_l/J_r}\quad,
 \label{eqn4}
 \end{equation}
 where
 \begin{equation}
 B_l\,=\;\frac{3(2 l^2 + 4l + 3)}{4 l \pi G \rho^2 R^2}\;=\;\frac{1}{e_g}\;\frac{4\pi(2l^2+ 4l + 3)}{3l}\quad,
 \label{eqn:Bl}
 \end{equation}
 while
 \begin{equation}
 e_g\,\equiv\;\frac{GM^2}{R^4}
 \label{eqn:eg}
 \end{equation}
 is (to order of magnitude) the gravitational energy density of the body. Here $\,J_r=1/\mu_r\,$ is the static (relaxed) compliance of the material, which is inverse to the static (relaxed) rigidity $\,\mu_r\,$. Switching from a static to an evolving configuration, we invoke the equivalence principle for viscoelastic materials, in order to obtain the complex Love number in the frequency domain:
 \begin{equation}
 \bar{k}_l(\chi) = \frac{3}{2(l-1)} \frac{1}{1+B_l/\bar{J}} = |\bar{k}_l(\chi)| e^{-i \epsilon_l(\chi)}\quad,
 \end{equation}
 $\,\chi \equiv |\omega|\,$ being the tidal frequency, and $\,\bar{J}(\chi)\,$ being the complex compliance of the material. Once 
 the compliance $\,\bar{J}(\chi)\,$ is prescribed by a rheological model, we can compute the quantity function
 \begin{equation}
 {k}_l(\chi)\;\sin \epsilon_l(\chi)\;=\;-\;{\rm Im} [\bar{k}_l(\chi)]\quad,
 \end{equation}
 where $\,{k}_l(\chi)\,\equiv\,|\,\bar{k}_l(\chi)\,|\,$ and $\,\epsilon_l(\chi)\,$ are, correspondingly, the degree-${\emph{l}}\,$ dynamical Love number and phase lag (the latter being linked to the degree-${\emph{l}}\,$ tidal quality factor through the equation \ref{L13}).

 However, an $\,lmpq\,$ term in the expansion (\ref{L10} - \ref{L11}) for the tidal torque includes the factor $\,{k}_l\,\sin \epsilon_l\,$ written not as a function of the tidal frequency $\,\chi \equiv |\omega|\,$ but as a function of the tidal mode $\,\omega\;\;$~---~see, e.g., the semidiurnal term given by the expression (\ref{eqn:T_simple}). It can be demonstrated that this brings an extra factor equal to the sign of the mode:
 \ba
 \nonumber
 [k_l  \sin \epsilon_l] (\omega)\;=\qquad\qquad\qquad\qquad\qquad\qquad\qquad\qquad~\qquad\\
  \label{eqn:klel}\\
 -\;\frac{3}{2(l-1)}\;
 \frac{B_l \; {\rm Im}({\bar J}(\chi))} {\left[\,  {\rm Re} (\,{\bar J} (\chi)\,) +  B_l \,\right]^2 + \left[\,{\rm Im} (\,{\bar J}(\chi)\,)\,\right]^2 }
 \times {\rm Sign}\,\omega~~.
 \nonumber
 \ea
 Whatever realistic rheological compliance $\,\bar{J}(\chi)\,$ is inserted into the above formula, the shape of the quality function is similar  for all viscoelastic bodies. It exhibits a sharp kink with two peaks having opposite signs (e.g., \citealt{noyelles14}, \citealt{efroimsky15}).$\,$\footnote{~A somewhat different approach to tides, explored by \citet{ferrazmello13, ferrazmello15a}, does not employ a constitutive equation explicitly. That model also predicts a similar shape for the quality function.}

 The generic shape of the quality function can be understood by comparing the frequency $\chi$ to the inverse
 of the viscoelastic relaxation time (e.g., \citealt{ferrazmello15b}.).
 At a fixed point in the body, the tidal stressing in the material is oscillating at the frequency $\,\chi \equiv |\omega|\,$. When $\,\chi\,$ is small compared to the inverse timescale of viscoelastic relaxation in the material, the body deformation stays almost exactly in phase with the tidal perturbation. The reaction and action being virtually in phase, no work is carried out and the tidal effects are minimal. On the other hand, at very high frequencies, the body's viscosity prevents it from deforming during the short forcing period $\,2\pi/\chi\,$. The reaction cannot catch up with the action, and stays close to zero~---~so, once again, little work is being done, and the tidal effects are again minimal.

 \subsection{The quality function for a Kelvin-Voigt sphere}
 
The goal of our paper is not to favour a particular rheological model, but to test our simulation method. 
The Kelvin-Voigt rheological model is simplistic,$\,$\footnote{~While it is still unknown what rheological models should describe comets and rubble-pile asteroids,  both seismic and geodetic data indicate that the Earth's mantle behaves viscoplastically as an Andrade body, see  \citet{efrolainey,efroimsky12a} and \citet{efroimsky12b}. At very low frequencies, its behaviour becomes Maxwell \citep{karato}.} but it is the easiest to model with a mass-spring model. 
Subsequent work will be aimed at extending our simulation approach to more realistic rheologies. 

\begin{figure}
    \includegraphics[width=3in]{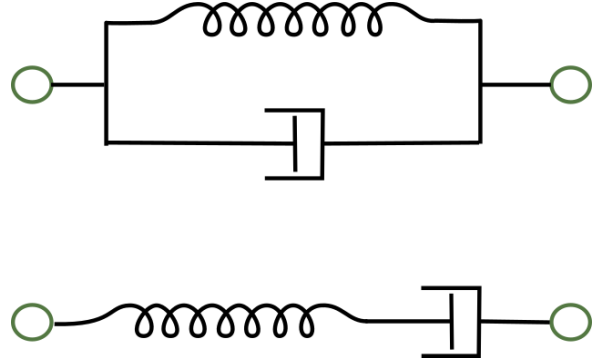}
 \caption{Green circles are mass elements.  In the Kelvin-Voigt model (the top drawing), spring and damping elements are set in parallel.
  In the Maxwell model (the bottom drawing), the two elements are arranged in series. Since it is easier to represent a Kelvin-Voigt viscoelastic material with a mass-spring network,
 we compute the quality function for the Kelvin-Voigt model, thus allowing a direct comparison between the quality functions calculated from theory and that computed through simulation.
 \label{fig:mass_spring}}
 \end{figure}

 The Kelvin-Voigt model of a viscoelastic solid can be represented by a purely viscous damper and a purely elastic spring connected to two mass elements in parallel. If we connect these two elements in series rather than in parallel, we describe the Maxwell model; Figure \ref{fig:mass_spring}. A Kelvin-Voigt body is easier to model with a mass-spring system, because both the spring and damping forces are directly applied to each node particle. This can  be seen from the  expression for the stress tensor $\,\sigma_{pq}\,$ as a function of the strain rate tensor $\,\dot{\varepsilon}_{pq}\,$ in the time domain (see \citealt{efroimsky12a}):
 \begin{equation}
 \sigma_{pq}(t) = 2 \int_{-\infty}^t \mu (t-t^{\prime}) \dot{\varepsilon}_{pq}(t^{\prime}) d t^{\prime}\quad,
 \end{equation}
 where $\,\mu (t-t^{\prime})\,$ is the stress-relaxation function. In the context of a mass-spring model, where the force between particles is likened to a uniaxial stress, the normal force applied to the particle $\,i\,$ due to a spring and dashpot 
 connecting it to particle $\,j\,$ is given by
 \begin{equation}
 F_i(t) = 2 \int_{-\infty}^t \mu (t-t^{\prime}) \dot{\varepsilon}(t^{\prime}) d t^{\prime}\quad.
 \label{eq1}
 \end{equation}
 For the Kelvin-Voigt model (e.g., \citealt{mase2010}),
\begin{equation}
\mu (t-t^{\prime}) = \mu + \eta \delta(t-t^{\prime})
\end{equation}
 where $\,\mu\,$ and $\,\eta\,$ are the unrelaxed shear rigidity and viscosity of the link between $\,i\,$ and $\,j\,$. Inserting that rheology into the equation (\ref{eq1}), we find
\begin{equation}
F_i(t) = 2 \mu \varepsilon(t) - 2 \mu \varepsilon(-\infty)  + 2 \eta \dot{\varepsilon}(t)\quad.
\end{equation}
 In neglect of the strain at $\,t=-\infty\,$, this becomes equivalent to  equations (\ref{eqn:elastic} - \ref{eqn:dampforce}) presented below and  employed in our code. 

 In mass-spring model simulations, massive particles are interlinked with a network of massless springs. To each of the two particles linked by a spring, a damping force is applied that depends on the spring strain rate (see Section 2.2 in \citealt{quillen15}). The shear elastic modulus, $\,\mu_I\,$, can be computed for an isotropic, initially random mass-spring model from the strength, lengths and distribution of the springs \citep{kot14}. We propose in the next section an estimate  for the simulated material shear viscosity $\,\eta_I\,$ based on the spring damping forces. Although the relation between the Kelvin-Voigt behavior of the damped springs, described above, and the bulk and shear properties of the material (i.e. the constitutive equations) is not obvious, our computations demonstrate that the simulated resolved body behaves similar to that expected for a Kelvin-Voigt solid with the actual elastic modulus $\,\mu \approx \mu_I\,$, viscosity $\,\eta\approx\eta_I\,$, and relaxation time
 \begin{equation}
 \tau\;=\;\frac{\eta}{\mu}\quad.
 \end{equation}
 There may be a difference between our $\,${\it{a priori}}$\,$ estimated values $\,\mu_I\,$, $\,\eta_I\,$ of the shear rigidity and viscosity (computed from the network), and the values $\,\mu\,$, $\,\eta\,$ of  the simulated material (see the discussion by \citealt{kot14}).

 We now derive the quality function (\ref{eqn:klel}) for the case of a Kelvin-Voigt sphere. We insert into the equation (\ref{eqn:klel}) the complex compliance of a Kelvin-Voigt body:
 \ba
 {\bar J}(\chi,\,\mu,\,\eta)\;\equiv\;\frac{\mu - i \chi \eta}{\mu^2 + \chi^2 \eta^2 }\;=\;\mu^{-1}\,\frac{1 - i \chi \tau}{1 + \chi^2 \tau^2}\quad.
 \label{eqKV}
 \ea
 The complex compliance $\,\bar J\,$ can be presented either as a function of the unrelaxed shear modulus $\,\mu\,$ and viscosity $\,\eta\,$, or as a function of the shear modulus $\,\mu\,$ and the relaxation time $\,\tau\,$. Introducing the dimensionless Fourier tidal mode
 \ba
 \tilde{\omega}\;\equiv\;\omega\;\tau \quad,
 \ea
 and following $\chi = |\omega|$, the dimensionless physical frequency
 \ba
 \tilde{\chi}\;\equiv\;\chi\;\tau\quad,
 \label{eqn:barchi}
 \ea
 we write down the complex compliance as
 \ba
 {\bar J}(\tilde{\chi},\,\mu)\;\equiv\;\mu^{-1}\;\frac{1 - i \tilde{\chi} }{1 + \tilde{\chi}^2 }\quad.
 \ea
 We define a dimensionless function
 \begin{eqnarray}
 {\bar j}^{(l)}({\tilde \chi},\mu ) & \equiv &  B_l^{-1} {\bar J}(\tilde \chi, \mu)\;=\;(B_l\,\mu)^{-1} \frac{1 - i \tilde \chi }{1 + \tilde \chi^2 }\quad,
 \end{eqnarray}
 with $\,B_l\,$ given by the expression (\ref{eqn:Bl}). Using that expression, we write down the function of the degree $\,l=2\,$ as
 \begin{eqnarray}
 {\bar j}^{(l=2)}(\tilde{\chi},\,\mu) &=&  \frac{3}{38 ~\pi}  \;\frac{e_g}{\mu}\;\frac{1\,-\,i\,\tilde{\chi}}{1\,+\,{\tilde{\chi}}^2 }\quad.
 \end{eqnarray}
 This is a function of the shear modulus $\,\mu\,$ given in the units of $\,e_g\,$, and of the frequency $\,\chi\,$ given in the units of the inverse relaxation time $\,\tau\,$. In terms of the dimensionless Fourier mode $\,\tilde{\omega}\,$ and the dimensionless frequency $\,\tilde{\chi}\equiv|\tilde{\omega}|\,$, the equation (\ref{eqn:klel}) becomes:
 \ba
 \nonumber
 [\,k_l \;\sin \epsilon_l\,] (\tilde{\omega},\;\mu)\;= \qquad\qquad\qquad\qquad\qquad\qquad\qquad\quad\quad\\
  \label{eqn:kl}\\
 -\;\frac{3}{2(l-1)}  \frac{{\rm Im}(\,{\bar j}^{(l)}(\tilde{\chi})\,)} {\left[\,{\rm Re} (\,{\bar j}^{(l)} (\tilde{\chi})\,) +  1\,\right]^2 + \left[\,{\rm Im} (\,{\bar j}^{(l)}(\tilde{\chi})\,)\,\right]^2 } \times
 {\rm Sign} \, \tilde{\omega}~~.
 \nonumber
 \ea
 This function attains its extrema at
 \begin{eqnarray*}
 \omega_{peak}^{(l)}\,=\;\pm\;\frac{\mu\;B_l\,+\;1}{B_l\;\eta}\;
 =\;\pm\;\frac{1}{\tau}\;\left( 1\;+\;\frac{1}{\mu\;B_l} \right)\quad.
 \end{eqnarray*}
 For $\,l=2\,$, the dimensionless peak mode and frequency are
 \begin{subequations}
 \begin{eqnarray}
 \tilde{\omega}_{peak}^{(l=2)}\,\equiv\,\omega_{peak}^{(l=2)}\,\tau\,=\,\pm\,\left(   1\,+\,  \frac{3}{38~\pi}  \;\frac{e_g}{\mu} \right)\qquad
 \end{eqnarray}
 and 
 \begin{eqnarray}
 \tilde{\chi}_{peak}^{(l=2)}\,\equiv\,\chi_{peak}^{(l=2)}\,\tau\,=\,|\omega_{peak}^{(l=2)}|\,\tau\,=\,\left(1\,+\,  \frac{3}{38 ~\pi}  \;\frac{e_g}{\mu}\right)~~.~~
 \end{eqnarray}
 \end{subequations}
 Elastic bodies should obey $\,\mu > e_g\,$, lest they collapse due to self-gravity.
 Hence we expect that
 \ba
 \omega_{peak}^{(l=2)}\;\tau\;\approx\;\pm\;1\quad.
 \ea
 Using the shorthand notation $\,\tilde{\omega}\,=\,\tilde{\omega}^{(l=2)}\,$ and
 \ba
 y(\mu, \tilde{\omega})\;\equiv\;  \frac{3}{38 ~\pi}  \; \frac{e_g}{\mu}\;(1\;+\;\tilde{\omega}^2)^{-1}\quad,
 \label{eqn:y}
 \ea
 the equation (\ref{eqn:kl}) for $\,l=2\,$ can be written as
 \begin{eqnarray}
 [k_2  \sin \epsilon_2] (\tilde{\omega},\;\mu)\;=\;\frac{3}{2}\;\frac{y\;\tilde{\omega}}{y^2\;(1 + \tilde{\omega}^2) + 2 y + 1}\quad.
 \label{eqn:pred}
 \end{eqnarray}
 We will use this expression for the quality function to analytical predict tidal response.

Classical tidal theory is valid only for incompressible materials (those having the Poisson ratio $\,\nu= 0.5\,$)~---~which is why the standard expression for the static Love number $\,k_2\,$ of an incompressible sphere depends only on the shear modulus of rigidity, not on the bulk modulus. However, the mass-spring models approximate a material with Poisson's ratio $\,\nu= 0.25\,$ \citep{kot14}. However, \citet{love11}  derived a general formula for the static Love number $\,k_2\,$ also for a $\,${\it{compressible}}$\,$ homogeneous sphere.$\,$\footnote{~Keep in mind that the assumptions of homogeneity and compressibility in Love's theory are mutually contradictive, wherefore this theory can be used only as an approximation \citep{melchior1972}.}  We have numerically checked that his formula for $\,k_2\,$ in the compressible case is only very weakly dependent on the Poisson's ratio, within broad ranges of the body size and density. This makes us confident that the standard tidal formulae (derived for the Poisson ratio $\,\nu= 0.5\,$) can  accurately describe a compressible body with Poisson ratio $\,\nu= 0.25\,$

\section{Damped mass-spring model simulations\label{sec3}}

We will compare the tidal spin down rate of a simulated viscoelastic body to that predicted analytically. To simulate tidal viscoelastic response we use the mass-spring model by \citet{quillen15}, that is based on the modular N-body code \texttt{rebound} \citep{rebound}. 
A random spring network model, rather than a lattice network model, was chosen so that the modeled body is approximately isotropic
and homogenous and lacks planes associated with crystalline symmetry.  We work in units of radius and mass of the tidally perturbed body: $\,R=1,\;M=1\,$. Time is given in units of $\,t_{grav} = \sqrt{R^3/GM\,}\,$ referred to as the $\,${\it{gravitational time scale}}. Pressure, energy density and elastic moduli are specified in units of $\,e_g\,\equiv\,GM^2/R^4\,$. In these units, the velocity of a massless particle on a circular orbit, which is just grazing the surface of the body, is $\,1\,$, the period of the grazing orbit being $\,2 \pi\,$.

\begin{figure}
 \includegraphics[width=3in]{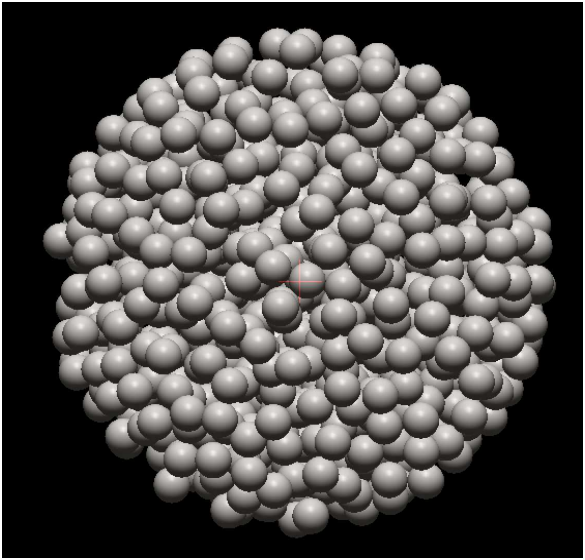}
 \caption{A snapshot of one of our simulations as viewed in the open-GL viewer of \texttt{rebound}.   We only show the tidally perturbed
 body, as the perturbing one is distant from it.  The rendered spheres are shown to illustrate the random distribution of node point masses,
 not imply that the body behaves as a rubble pile (e.g.,  \citealt{richardson2009, sanchez2011}).  Nodes are connected with a network of damped elastic springs.
 \label{fig:part}}
 \end{figure}

 Modeling the tidally perturbed body, we randomly generate an initial spherical distribution of particles of equal mass (as described in Section 2.2 in \citealt{quillen15}); see our Figure \ref{fig:part} for an illustration. 
 Particle positions are randomly generated using a uniform distribution in three dimensions 
 but a  particle is added (as a node) to the spring network only if it is sufficiently
separated from other particles (at a distance greater than minimum distance $d_{I}$) and  within a radius of $\,R = 1\,$ from the body centre.
Springs are added between two nodes if the distance between the nodes
is less than distance $d_s$. 

 The particles are subjected to three types of forces: the gravitational forces acting on every pair of particles in the body and with the massive companion, and the elastic and damping spring forces acting only between sufficiently close particle pairs. Springs with a spring constant $\,k_I\,$ interconnect each pair of particles closer than some distance $\,d_s\,$. The rest length of each spring is initially set equal to its initial length. The springs experience compression or extension relative to their rest length. The number of springs and their rest lengths stay fixed during our simulations. The springs have different rest lengths, and we denote their average initial (rest) length with $\,L_I\,$. The total number of particles is $\,N_I\,$, and the total number of springs is $\,NS_I\,$. The mass of each particle is $\,m_I = 1.0/N_I\,$,  and the initial mass density is approximately uniform.

The number and distribution of particles and links in a mass-spring model is mainly set by the minimum distances $d_I, d_s$, however
because the initial particle positions are randomly generated, there are local variations in density of nodes and the spring network.
The ratio $d_s/d_I$ sets the mean number of springs per mass node.

Consider a spring linking the particle $\,i\,$ residing at $\,{\bf x}_i$ with a particle $\,j\,$ located in $\,{\bf x}_j\,$. The force due to the spring on each mass node is computed as follows. The vector pointing from one of these particles to another, $\,{\bf x}_i - {\bf x}_j\,$, gives the spring length $\,L_{ij} = |{\bf x}_i - {\bf x}_j|\,$ that we compare with the spring rest length $\,L_{ij,0}\,$. The elastic force exerted upon on particle $\,i\,$ by the particle $\,j\,$ is 
 \begin{equation}
 {\bf F}_{i}^{elastic} = -k_I (L_{ij} - L_{ij,0}) \hat {\bf n}_{ij}
 \label{eqn:elastic}
 \end{equation}
 where  $\,k_I\,$ is the spring constant, while the unit vector is $\,\hat {\bf n}_{ij} = ( {\bf x}_i - {\bf x}_j )/ L_{ij}\,$.
 The appropriate force acting on the particle $\,j\,$ from the particle $\,i\,$ is equal in magnitude and opposite in direction.

The strain rate of a spring with length $L_{ij}$ is
\begin{equation}
 \dot \epsilon_{ij} = \frac{\dot L_{ij}}{L_{ij,0}} = \frac{1}{L_{ij} L_{ij,0}} ({\bf x}_i - {\bf x_j}) \cdot ({\bf v}_i -  {\bf v}_j)
 \end{equation}
 where ${\bf v}_i$ and ${\bf v}_j$ are the particle velocities and  $\dot L_{ij} $ is the rate of change of the spring length.
 To the elastic force acting on the particle $\,i\,$, we add a damping (viscous) force proportional to the strain rate:
 \begin{equation}
 {\bf F}_{i}^{damping} = - \gamma_I  \dot \epsilon_{ij} L_{ij,0} m_I \hat{\textbf{n}}_{ij}\quad,
 \label{eqn:dampforce}
 \end{equation}
 with a damping coefficient $\,\gamma_I\,$  equal to the inverse damping  time scale. The parameter $\,\gamma_I\,$  is independent of $\,k_I\,$.
 As all our particles have the same mass, we do not use the reduced mass in equation (\ref{eqn:dampforce}), as did \citet{quillen15}.

 How does the spring constant $\,k_I\,$ and the damping parameter $\,\gamma_I\,$ relate to the global rigidity and viscosity of the body? According to \citet{kot14}, the static Young's modulus is given by a sum over the springs $L_{ij,0}$:
 \begin{equation}
  E_I\,=\;\frac{1}{6V}\;\sum k_I L_{ij,0}^2\quad,
 \label{eqn:Emush}
 \end{equation}
 where $\,V\,$ is the total volume. For an initially random isotropic mass-spring system, the Poisson ratio is $\,\nu = 0.25\,$ \citep{kot14}. Our mass-spring model allows us to directly estimate the Young's modulus $\,E_I\,$, from which we can compute the shear elastic rigidity $\,\mu_I\,$ commonly used for tidal evolution calculations. The relation between the two relaxed (static) moduli is given by
 \begin{equation}
 \mu_I = \frac{E_I}{2(1+\nu)} = \frac{E_I}{2.5}\quad,
 \label{eqn:mu_I}
 \end{equation}
 where we set the Poisson ratio to be $\,\nu=0.25\,$.

 With non-zero damping coefficients, the stress is a sum of an elastic term proportional to the strain and a viscous term  proportional to the strain rate, so the model should locally approximate a linear Kelvin-Voigt rheology. The mass-spring model is compressible. So, when damped springs are used, it exhibits a bulk viscosity (in analogy to the bulk modulus) and a shear viscosity (in analogy to the shear modulus). For a mass-spring model comprised of equal masses $\,m_I\,$ and parameterised with the damping coefficients $\,\gamma_I\,$ and spring constants $\,k_I\,$, we can tentatively estimate the shear viscosity $\,\eta_I\,$ as the ratio of the damping and elastic forces given by the equations (\ref{eqn:dampforce}) and (\ref{eqn:elastic}), correspondingly. We also assume that $\,\eta_I\,$ scales with the network in the same way $\,\mu_I$ does in equations (\ref{eqn:mu_I}) and   (\ref{eqn:Emush}). The resulting estimate for the viscosity is:
 \begin{equation}
 \eta_I \approx \mu_I \left( \frac{\gamma_I m_I}{k_I}\right)\quad.
 \label{eqn:eta}
 \end{equation}
 The relaxation time scale of a Kelvin-Voigt solid is
 \begin{equation}
 \tau_{relax}~=\;\frac{\eta_I}{\mu_I}\;\approx\;\frac{\gamma_I\;m_I}{k_I}\quad.
 \label{eqn:taurelax}
 \end{equation}

 While the fidelity of the computed elastic modulus of a mass-spring model has been checked by numerical simulations using static
 applied forces \citep{kot14}, the viscosity of a damped mass-spring model has not been tested. We consider the equations (\ref{eqn:eta}) and (\ref{eqn:taurelax}) as approximate. They may need to be amended with factors of order unity.
 We shall discuss this possibility later, when we compare measurements from our
 simulations to predictions from an analytical model of tidal evolution.

 In our simulations, the body was given an initial spin $\,\dot{\theta}=\sigma_0\,$ perpendicular to the orbital plane. 
 This was done by setting the initial velocities of particle equal to
 \begin{equation}
 {\bf v}_i\,=\;{\bf x}_i\,\times\,\sigma_0\,\hat {\bf z}\quad,
 \end{equation}
 $\,{\bf v}_i\,$ and $\,{\bf x}_i\,$ being the velocity and position vectors of the $\,i\,$-th particle, with respect to the body's centre of mass,
 and the unit vector $\,\hat {\bf z}\,$ being orthogonal to the orbit. In most runs, we chose the initial rotation to be retrograde ($\,\sigma_0<0\,$), to allow for a larger range of values of the tidal frequency $\,\chi\,$ to be simulated. This choice always led to a decrease in the semimajor axis (see Table \ref{tab:list}) and to acceleration of the body's spin rate. Variations in the initial particle distribution creates only negligible non-diagonal terms in the inertia tensor in this coordinate system.

 From the \texttt{rebound} code, version 2 as of November 2015, we used the open-GL display with open boundary conditions, the direct all-pairs gravitational force computation, and the leap-frog integrator needed to advance particle positions. To the particles' accelerations caused by the gravity, we added the additional spring and spring-damping
 forces, as was explained above. To maintain numerical stability, the time step was chosen to be smaller than the elastic oscillation frequency of a single node particle in the spring network, 
 or equivalently the time it takes vibrational waves to travel between two neighbouring nodes.

\section{Tidal evolution of the semimajor axis\label{sec4}}

\subsection{The semimajor axis' tidal drift rate computed from the mass-spring model\label{4.1}}

 Common parameters for our first set of simulations are listed in Table \ref{tab:common}, along with their chosen or computed values. 
 A list of varied and measured parameters is presented in Table \ref{tab:var}. The values used in different individual 
 simulations with common values in Table \ref{tab:common} are listed in Table \ref{tab:list}.
 We chose the values for the mass ratio $\,M^*/M\,$ and the initial semi-major axis $\,a_0\,$ to be the same in the two sets of simulations performed. However, in each set, the spring damping rate $\,\gamma_I\,$ and the initial body spin rate $\,\sigma_0\,$ were chosen to sample a range of values of the Fourier tidal mode frequency $\,\omega\,$ and of the viscoelastic relaxation time $\,\tau_{relax}\,$. Using the equation (\ref{eqn:Emush}), we computed the Young's modulus by only considering the springs with a midpoint radii less than $0.9R$, and we used the volume within the same radius $\,0.9R\,$.    The region near the surface was discarded so 
 that the Young's modulus was computed in a region where the spring network is isotropic.
 A fairly soft body under strong tidal forcing (corresponding to a large mass ratio and weak springs) was chosen, to reduce the integration time required to observe a significant tidally induced change in the semimajor axis. At the same time, we made sure that the Young's modulus was sufficiently large, so that the body was strong enough to maintain a nearly constant radial density profile. In the absence of exterior pressure, the body is held up against self-gravity by spring forces only~---~so the springs in the interior are under compression. In all our runs, the particles were displaced, in the body frame, by at most a few percents of the unit length $\,R\,$.

 We chose the initial semi-major axis $\,a_0\,$ large enough, to ensure that the quadrupole tidal potential term would dominate. We also set the initial relative velocity of the bodies such that the orbit would be circular. In the frame corotating with the tidally perturbed body, the perturber orbits with a period of $\,P_o\,=\,2\,\pi/\chi\,$.
 We carried out each run over the time span of $\,t\,=\,11\,P_o\,$ and recorded the semimajor axis' values 10 times, with even intervals of $\,P_o\,$. Since the springs' lengths had initially been set to have their rest values, gravity caused the system to bounce at the beginning of each simulation. Damped oscillations are expected in our model as  each of the individual links between pairs of particles is acting as a damped harmonic oscillator. The  initial oscillations are just the consequences of abruptly ``turning on'' self-gravity at the beginning of the simulations. Thus, during the first time interval of $\,P_o\,$, we integrated the body with a higher damping parameter,  to dissipate the initial vibrational oscillations. We did not take into account the semimajor axis' value computed during that time interval. Subsequently, we recorded the semi-major axis' values at an interval of $\,P_o\,$, so that the irregularities of the particle distribution did not affect the measurement of the slowly drifting semi-major axis. 

 The semimajor axis was computed using the distance between the centre of mass of the bodies, and their relative velocities. To measure the semimajor axis' drift rate $\,\dot a\,$, we fit a line to the ten measurements of the semimajor axis at the ten time intervals $\,P_o\,$. ~Each fit was individually inspected, to ensure that the ten points lay on a line. The standard error of the fitted slope value that provides $\,\dot a\,$  was $\,\lesssim 1\%\,$. For each simulation, we also compared the initial and final values of each component of the total angular-momentum vector (relative to the centre of mass of the binary), and found the absolute difference was below $\,10^{-12}\,$ for all the components. Similarly, we checked that the evolution of the measured spin rate and semimajor axis were tightly anti-correlated, as expected when angular momentum is conserved. The total energy was not conserved because of the spring damping forces.
 
 For each simulation we generate a new mass-spring network.  Hence each simulation has slightly different numbers of mass nodes
 and springs and variations in the node distribution and associated spring network.   
Two simulations run with identical input parameters will differ slightly in their measured semi-major axis drift rates.
We set the minimum distance between nodes $d_I$ and maximum spring length $d_s$ (setting the mean number of mass nodes
and numbers of springs) 
so that the differences in semi-major axis drift rate for simulations drawn from the same parameter set differed by less than 10\%.
We will discuss the sensitivity of the simulations to the numbers
of nodes and springs per node further below.

The simulations were run on a MacBook Pro (early 2015) with  3.1 GHz Intel Core i7 microprocessor.
The computation time for an individual simulation listed in Tables \ref{tab:common} and \ref{tab:list} was about 7 minutes.
If the number of mass nodes is doubled then the number of direct all-pairs gravity force computations increases by 
a factor of 4.
However the total computation time increases by a slightly larger  factor than 4 because the  number of springs
also increases (scales with the number of nodes) and the timestep decreases.
For simulations with similar vibration wave speeds, the timestep scales with the interparticle spacing
and so the total number of particles to the -1/3 power.

\subsection{Comparison of numerically measured and predicted quality functions\label{4.2}}

\begin{figure*}
   \includegraphics[width=3.5in]{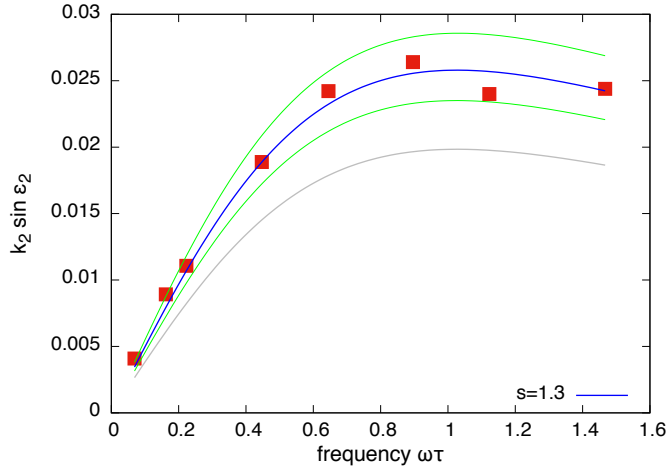}
 \caption{A comparison of the numerically computed quality function, shown as points, to that calculated analytically for a homogeneous Kelvin-Voigt sphere.
 The analytically derived frequency-dependence is given by the grey curve (obtained with equation \ref{eqn:pred}), and when
 multiplied by $s$, in blue, with the scaling factor $s$ listed on bottom right. The green curves show the effect of raising and lowering the shear modulus by 10\% in the offset analytical calculation. Both the numerical and analytical calculations were carried out for a perturber of mass $\,M^*=100\,$ and initial value of the semi-major axis  $\,a_0=10\,$. We find that the numerically computed quality function  has a shape and peak frequency consistent with that obtained analytically, but the amplitude is too high by about 30\%.  
 We attribute the scatter of the points off the line to variations in the value
 of the shear modulus of the random mass-spring network due to non-uniformity 
 in the particle distribution and spring network (see section \ref{4.2}).
   \label{fig:ff}}
   \end{figure*}

\begin{figure*}
   \includegraphics[width=3.5in]{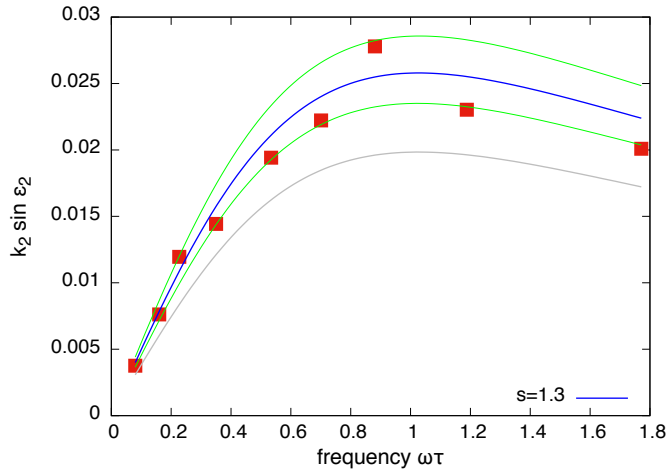}
   \caption{The same as in Figure \ref{fig:ff}, except that the perturber's mass is $\,M^*=200\,$, and the initial value of the semi-major axis is $\,a_0=100\,$.
   \label{fig:ff2}}
   \end{figure*}

 Inverting the equation (\ref{eqn:daa}), we obtain the following expression for the quality function:
 \begin{equation}
 [k_2 \sin \epsilon_2](\tilde \chi)\;= \;-\;\frac{\dot a}{3na}
 \left( \frac{M}{M^*} \right)\left( \frac{a}{R} \right)^5 \quad.
 \label{eqn:qual}
 \end{equation}
 From the semimajor axis' drift rates computed numerically in our simulations, we computed the values of the quality function over a range of values of frequency. This was performed  using the equation (\ref{eqn:qual}) and the quantities listed in Table \ref{tab:list}. The  numerically measured values of the quality function are plotted as a function of the dimensionless frequency $\,\tilde \chi\,$ in Figures \ref{fig:ff} and \ref{fig:ff2}. The dimensionless frequency $\bar \chi$ was computed by using the relaxation time that was evaluated for each simulation individually.

 The numerically generated points in Figure \ref{fig:ff} show that the simulated quality function is proportional to the frequency, at small frequencies, but decays at large frequencies. Predicted analytically for various rheologies \citep{efroimsky12a, noyelles14, efroimsky15}, this behaviour has never been reproduced by direct numerical computations. The points used to build this plot were measured from simulations with different spin rates and relaxation times. Nevertheless, they all lie near the same curve, suggesting that the function is primarily dependent on the relaxation time. We thus have numerically confirmed the expected behaviour and sensitivity of the quality function  to the frequency and to the viscoelastic timescale.

With the points obtained through simulations, we plot the quality function predicted analytically using equation (\ref{eqn:pred}). 
We set the shear modulus $\,\mu = \mu_I\,$, with the value of $\,\mu_I\,$ given by the expression (\ref{eqn:mu_I}). This value is proportional to the Young's modulus listed in Table \ref{tab:common} computed from the spring network. The mass ratio and mean motion are taken from Table \ref{tab:list}. The result is the grey line (the lowest one) in Figure \ref{fig:ff}. This is the analytically predicted quality function, and we see that its numerically obtained counterpart (given by the red points) is higher. We refer to the ratio of numerically measured quality function to the predicted one as $qf_{ratio}$ and from this plot we find $qf_{ratio}\sim 1.3$.

 Figure \ref{fig:ff} demonstrates that the numerically obtained tidal drift rate of the semimajor axis is maximal at a frequency $\,\tilde{\chi} \approx 1.0\,$, which is consistent with the analytical model. Had the relaxation timescale (equation \ref{eqn:taurelax}) been miscomputed in our simulations, the numerically obtained peak would have been displaced away from that predicted analytically. The  good match between predicted and numerically measured peak frequency supports our estimate for the shear viscosity (equation \ref{eqn:eta}) and the associated relaxation time (equation \ref{eqn:taurelax}) for the mass-spring model.

 Both the peak magnitude and the peak location of the quality function may shift if higher-order terms (with higher values of $\,l\,$) are included into the analytical calculation.
 As this might explain the difference in height of our numerically computed quality function, compared to that predicted, we tested this possibility by running simulations with a larger value of the semimajor axis. Simulations with a larger initial semimajor axis are also listed in Table \ref{tab:list}, and the quality function for these runs is plotted in Figure \ref{fig:ff2}.
 We find that the amplitude correction factor required to match the numerical results is the same as for the previous set of simulations~---~and, again, the frequency does not need to be rescaled.
 From this, we conclude that higher-order terms in the tidal potential do not explain the amplitude discrepancy between our numerical simulations model and our analytical predictions.

 Since we are using a random mass-spring model, both the particle distribution and the spring network differ between simulations.  We computed the Young's modulus and relaxation time for each run, and these are listed in Table \ref{tab:list}. We found small variations in the elastic modulus between different simulations. In Figures \ref{fig:ff} and \ref{fig:ff2}, we also show the analytically predicted quality function factored by 1.3 and offset by $\,\pm\,10\,\%\,$ (green curves). Points with higher values of $\,E_I\,$, corresponding to harder bodies, systematically lie lower than the appropriate points with lower values of $\,E_I\,$ corresponding to softer bodies  experiencing stronger tidal deformation. The scatter in our points above and below the blue line can  be attributed to variations in the particle distribution and in the associated spring network.

\begin{table}
\vbox to70mm{\vfil
\caption{\large  Common simulation parameters   \label{tab:common}}
\begin{tabular}{@{}lllllll}
\hline
\hline
$N_I$ & 800 & Number of particles in resolved body \\
$\NS_I$ & 9254 & Number of interconnecting springs  \\
$L_I$  & 0.2623& Mean rest spring length \\
$k_I$  & 0.08 & Spring constant  \\
$E_I$  & 2.3 & Mean Young's modulus  \\
$d_I$    &0.15 &  Minimum initial interparticle distance \\
$d_s$ & 0.345 &  Spring formation distance \\
$dt$   &0.001 & timestep \\
\hline
\end{tabular}
{\\ $N_I$,  $\NS_I$,  $E_I$ and $L_I$ vary slightly between simulations as particle distributions are randomly generated.
$E_I$ is computed using equation (\ref{eqn:Emush}) and is the average value for all the simulations.
These parameters are common to simulations listed in Table \ref{tab:list}.
}
}
\end{table}

\begin{table}
\vbox to 70mm{\vfil
\caption{\large Description of varied and measured simulation parameters   \label{tab:var}}
\begin{tabular}{@{}lllllll}
\hline
$a_0$    & initial semi-major axis \\
$M^*/M$   & mass ratio  \\
$ \tilde \chi$  & Unitless frequency \\
$\dot a $ & Rate of orbital decay in semi-major axis \\
$\gamma_I$ & Spring relaxation time \\
$ \sigma_0$  & Initial spin \\
$\tau_{relax}$ & Estimated relaxation time of viscoelastic solid  \\
$\chi $  & Initial tidal frequency (semi-diurnal) \\
$P_o$   & Time between integration outputs \\
$E_I$ & Computed Young's modulus \\
\hline
\end{tabular}
{\\ The viscoelastic relaxation time $\tau_{relax}$ is computed using the equations (\ref{eqn:Emush}) and (\ref{eqn:taurelax}).
 The frequency $\,\chi\,$ is defined by the expression (\ref{eqn:omega}), while the tidal forcing frequency $\tilde \chi$ is given by the equation (\ref{eqn:barchi}).
 The period $\,P_o = 2\pi/\chi\,$ is also a part of the simulation output.
}
}
\end{table}

\begin{table}
\vbox to 120mm{\vfil
\caption{\large  Quantities either set or computed in different mass-spring N-body simulations  \label{tab:list}}
\begin{tabular}{@{}lrrrrrrrr}
\hline
 $ \tilde \chi$  & $\dot a $&$\gamma_I$ & $\sigma_0$  &  $\tau_{relax}$ &  $\chi $  &  $P_o$ & $E_I$ \\
\hline
\multicolumn{5}{l} {with $a_0 = 10$, $M^* = 100$, $n=0.318$} &&& \\
\hline
 0.069 & -3.897e-05 & 20 & 0.1 & 0.158 & 0.436 & 14.424 & 2.30   \\
0.162 & -8.505e-05 & 20 & -0.2 & 0.156 & 1.036 & 6.067 & 2.35    \\
0.223 & -1.055e-04 & 20 & -0.4 & 0.156 & 1.436 & 4.377 & 2.40    \\
0.447 & -1.800e-04 & 40 & -0.4 & 0.311 & 1.436 & 4.377 & 2.38    \\
0.645 & -2.309e-04 & 50 & -0.5 & 0.394 & 1.636 & 3.841 & 2.33   \\
0.896 & -2.517e-04 & 80 & -0.4 & 0.624 & 1.436 & 4.377 & 2.36    \\
1.123 & -2.288e-04 & 100 & -0.4 & 0.782 & 1.436 & 4.377 & 2.36    \\
1.467 & -2.325e-04 & 100 & -0.6 & 0.799 & 1.836 & 3.423 & 2.26    \\
\hline
\multicolumn{5}{l}  {with $a_0 = 20$, $M^* = 200$, $n=0.159$ }&&& \\
\hline
0.080 & -2.239e-06 & 20 & -0.1 & 0.156 & 0.517 & 12.153 & 2.40   \\
0.160 & -4.540e-06 & 40 & -0.1 & 0.309 & 0.517 & 12.153 & 2.41   \\
0.227 & -7.124e-06 & 40 & -0.2 & 0.317 & 0.717 & 8.763 & 2.29    \\
0.350 & -8.601e-06 & 40 & -0.4 & 0.314 & 1.117 & 5.625 & 2.36    \\
0.534 & -1.157e-05 & 60 & -0.4 & 0.478 & 1.117 & 5.625 & 2.25  \\
0.701 & -1.324e-05 & 80 & -0.4 & 0.627 & 1.117 & 5.625 & 2.26    \\
0.881 & -1.656e-05 & 100 & -0.4 & 0.789 & 1.117 & 5.625 & 2.28    \\
1.188 & -1.372e-05 & 100 & -0.6 & 0.783 & 1.517 & 4.142 & 2.31    \\
1.771 & -1.197e-05 & 150 & -0.6 & 1.167 & 1.517 & 4.142 & 2.28   \\
\hline
\end{tabular}
{\\  The rightmost column contains the values of the Young's modulus, computed for each simulation by means of the equation (\ref{eqn:Emush}).  The simulations listed here have common parameters listed in Table \ref{tab:common}.
}
}
\end{table}

\subsection{Discrepancy in amplitude of quality function and sensitivity to the number of masses and springs simulated \label{4.3}}
 
To see how the  discrepancy in amplitude is related to the number of mass nodes and springs,
 we carried out a second series of simulations each with   the same 
 initial spin, estimated elastic modulus and viscoelastic relaxation timescale, but having different numbers of node masses and 
 numbers of springs per node.  Parameters for these simulations are listed  in  Table
 \ref{tab:NN}.  We ran 6 sets, each of 5 simulations, all approximately matching the third row in Table \ref{tab:list} with
 $a_0=  10$ , $M_*/M = 100  $,  $\sigma_0= -0.4 $,  and $P_o =  4.377 $.  
 The 5 simulations in each set have identical run parameters.
 The first 10 simulations (columns A, B in Table \ref{tab:NN}) have about 400 mass nodes, 
 the second 10 (columns C, D)  about 800 mass nodes and the last 10 (columns E, F) about 1600 mass nodes.
 The first 5 in each group of 10 have about 10 springs per node (columns A, C and E) whereas the second 5 of each group of 10 (columns B, D and F)    have about 20 springs per node.
 The spring constants, $k_I$, and damping parameters, $\gamma_I$, for each set were chosen so that $\bar \chi \approx 0.225$,  
 $E_I \approx 2.3$ and $\tau_{relax} \approx 0.155$.
 For each simulation separately we computed ratio $qf_{ratio}$ of the numerically measured value of the quality function $k_2 \sin \epsilon_2$
 to the predicted one computed using the normalized frequency $\bar \chi$ and Young's modulus measured from the 
 nodes and springs in the simulations.    We then computed the mean and standard deviation of $qf_{ratio}$
 for each  set of  5 simulations with identical run parameters and these are listed
in the bottom  rows of  Table \ref{tab:NN}. 
 The standard deviation in semi-major axis drift rate divided by the mean value of the drift rate
  is also computed for each set of 5 simulations and listed
 in  Table \ref{tab:NN}. 
 
 In the simulations listed in Table \ref{tab:common} and \ref{tab:list}, the ratio of the number of springs per node is about 11, which is slightly lower than the number recommended by \citet[Figure 5]{kot14}. For this ratio, \citet{kot14} found that the spring network behaved 10\% weaker~\footnote{~Because of a misprint, Figure 5 in \cite{kot14} actually displays $\,(E_0-E)/E_0\,$ instead of $\,(E-E_0)/E_0\,$, with $\,E_0\,$ and $\,E\,$ being the estimated and measured Young's modulus, respectively (Kot et al., private communication). A smaller average number $\,\langle S \rangle\,$ of springs per node gives a smaller Young's modulus and a weaker body.} than what was estimated by  equation (\ref{eqn:Emush}).   In Table \ref{tab:NN} we can compare the ratio of numerically measured to
 predicted quality functions for simulations with similar numbers of nodes but differ by the number
 of springs per node.   
 Comparing columns A to B and C to D and E to F we see that the ratio of numerical measured to predicted
 quality function is only slightly less (about 2\% lower) when the number of springs per node is about 20 rather than 10.
 We find that for greater than 10 springs per nodes, the ratio of measured to predicted quality function is relatively insensitive to the number of springs per node.

As each simulation generates a new particle distribution and spring network, we can measure effects due to variations in these properties
by comparing simulations generated from identical input parameters.   Table \ref{tab:NN} shows that the standard deviation of the quality function ratio and semi-major axis drift rate decreases with increased particle number
and that the scatter in these measured quantities differs by only a few percent.
We conclude that variations in the spring network are unlikely to explain the discrepancy between predicted
and numerically measured quality function.

 One possible reason for the discrepancy between predicted and measured quality function is that near the body surface the number of springs per particle is lower  than the interior, and thus the spring network is anisotropic near the surface. The simulated body is weaker (and floppier) than we estimated by integrating the spring properties over the volume.  Because the overall rigidity of the body is weaker, its tidal response would be larger than we predicted analytically.    This would be consistent with our greater than unity $q f_{ratio}$. 
   If this were the primary cause of our amplitude 
 discrepancy then the ratio of numerical computed to analytical predicted quality function should inversely depend on the number
 of simulated masses.   Table \ref{tab:NN} shows that this ratio $qf_{ratio}$ does decrease as the particle number is
 increased from 400 to 1600, however it only decreases by about a percent between 800 and 1600 particles.
 Convergence has not been achieved, but the standard deviations in quantities measured from groups of simulations
imply that the simulation results are reproducible and that the scatter is smaller than the amplitude discrepancy.  
If doubling the particle number reduces the quality function
ratio by 2\% then to reach a quality function ratio of 1 we would require a million particles in the simulation
and we are not yet set up to run simulations this large.

The surface of our body lies within a radius of 1, consequently our simulated body effectively has an outer radius
that is smaller than 1.
As the semi-major axis drift rate is faster for larger bodies, we would have expected our simulations
to exhibit slower rather than faster orbital decay rates (and evident from the factor of $R^{-5}$ in equation \ref{eqn:qual}).   
We recall that the analytically predicted quality function depends on the ratio $\mu/e_g$ (see equations \ref{eqn:y}
and equation \ref{eqn:pred}).   As the energy density $e_g \propto R^4$, for an effective radius less than 1, the energy
density is higher than we previously estimated and the ratio $\mu/e_g$ would be lower than we used.
This means we have underestimated the size of the tidal response.    This is in the right direction to
account for the amplitude discrepancy.  However if we take into account both the factor of $R^5$ in equation \ref{eqn:qual}
when computing our numerical quality function 
and the factor of $R^4$ in equation \ref{eqn:pred} when computing our analytical quality function, the two corrections
more or less cancel each other and we do not resolve the source of discrepancy in quality function amplitude.
So we find that we cannot resolve our discrepancy in amplitude
by correcting the body radius to an effectively smaller radius.

 Equations (\ref{eqn:Emush} - \ref{eqn:mu_I}) characterise static properties of the mass-spring model.  However, the analytical theory of evolving tides (including that based on the viscoelastic rheology \ref{eqKV}) employs the unrelaxed shear rigidity $\,\mu\,$. We have assumed that  the two values are equivalent, although this is  not perfectly true for real materials. As any rheological parameter, the shear elasticity modulus depends upon frequency.    The unrelaxed or frequency dependent shear modulus in real materials can be lower 
 than the relaxed or static counterpart (e.g., \citealt{fj2005}).  Perhaps this is also true in our simulated material.  The numerical
 tests by \citet{kot14} used  static forces and so would not have been sensitive to frequency dependence in the shear modulus.
 Had we employed  a smaller value for the shear modulus, this would have increased the values of the theoretically obtained $\;k_2\,\sin \epsilon_2\;$, and thus would have reduced the offset between our numerical and theoretical values for the quality function.   

 Each numerical run starts out with a sphere of particles, with springs at their rest lengths. However, after the simulation begins, the body is compressed by self-gravity. The resulting density profile is not perfectly flat, as the centre becomes more compressed than the regions closer to the surface. Furthermore, the initial spin of the body causes its shape to deviate from a sphere. To sample a broad range of frequencies (in units of the relaxation time), and to have a stronger tidal response (i.e., to reduce the simulation time), we use a soft body that is particularly prone to deformation when spun and compressed by self-gravity.   Nevertheless our simulated bodies are not
 strongly deformed and we do not expect body deformation and compression to account for our quality function amplitude discrepancy.
 
 Since our simulated body is comprised of randomly-distributed point particles, the body is neither exactly spherical nor uniform. Initially, its principal axes of inertia (computed from the moment of inertia tensor) are oriented randomly, and the three moments of inertia are not exactly equal. So, initially, the body's spin angular momentum is not exactly perpendicular to the orbit. In the course of the simulations, we measured the values of the $\,x\,$ and $\,y\,$ components of the spin angular momentum,  finding them to be a few hundredths of the initial $\,z\,$ component.  This is small enough that  the spin about non-polar axes is unlikely to be the cause of our quality function amplitude discrepancy.
 
 To summarize, we have compared simulations with different numbers of mass nodes and springs per node and found
 that the scatter and sensitivity of the ratio of numerically measured to analytically predicted quality function
 are only weakly dependent on these quantities.  Convergence has not been reached at 1600 simulated particles
 but the scatter
 in the simulations is low enough that the variations in the spring network are small enough that we have
 reproducible numerical measurements and scatter within these measurement are smaller than
 our measured amplitude discrepancy.   We have not identified the source of the 30\% discrepancy
 in the amplitude of our numerically measured quality function, however we suspect variations in the spring network, 
 floppiness in the outer surface and a possible frequency dependence in the behavior of the numerically simulated
 shear modulus.

\begin{table}
\vbox to 120mm{\vfil
\caption{\large  Comparison of simulations with different numbers of node masses and springs per node  \label{tab:NN}}
\begin{tabular}{@{}lrrrrrrrr}
\hline
\multicolumn{7}{l} {with $a_0 = 10$, $M^* = 100$, $n=0.318$ , $\sigma_0= -0.4$,  $P_o =  4.377$ }  \\
\multicolumn{7}{l} {and $\bar \chi \sim 0.225$,  $E_I \sim 2.3$, $\tau_{relax} \sim 0.155$ }  \\
\hline
Set                           &  A     &  B    &   C   &   D   &   E    &  F    \\
$\langle N_I  \rangle$    & 403.8  & 405.8 & 795.2 & 804.2 & 1597.8 & 1596.2 \\
$\langle NS_I/N_I\rangle$ & 10.92  & 18.7  & 11.5  & 20.34 & 12.02  & 21.34 \\
$d_I$                     &  0.19  & 0.19  & 0.15  & 0.15  & 0.118  & 0.118 \\
$d_s/d_I$                 &  2.3   & 2.8   & 2.2   & 2.8   & 2.3    & 2.8   \\
$k_I$                     & 0.102  & 0.04  & 0.08  & 0.0301& 0.062  & 0.023 \\
$\gamma_I$                & 12.858 & 5.05  & 20.0  & 7.48  & 31.3   & 11.5  \\
$ \langle qf_{ratio} \rangle$           & 1.48   & 1.39  & 1.41  & 1.36  & 1.40   & 1.38  \\
$\sigma[ qf_{ratio} ]$              & 0.069  & 0.052 & 0.053 & 0.026 & 0.030  & 0.012 \\
$\sigma [\dot a]/      \langle \dot a \rangle  $                     & 0.061 &0.064 &0.055 &0.024& 0.026& 0.0094 \\
\hline
\end{tabular}
{\\ Each column represents a group of 5 simulations.   From each group of 5 the mean number of nodes and springs per node
are listed in the second and third rows.   
The  rows labelled $ \langle qf_{ratio} \rangle$ and $\sigma[ qf_{ratio} ]$ give the mean and standard deviations, respectively, of the
ratio of the numerically measured to analytically predicted
quality function. The bottom row gives the ratio of the standard deviation in the semi-major axis drift rate divided by the means.
Each of the means and standard deviations are computed from 5 simulations.
The simulations listed here differ from those described by Tables \ref{tab:common} and \ref{tab:list}.
}
}
\end{table}

\section{Conclusion\label{sec5}}

 In this article, we have used a self-gravitating damped mass-spring model, within an N-body simulation, to directly model the tidal orbital evolution and rotational spin-up of a viscoelastic body.

 We considered a binary comprised of an extended body (assumed spherical and homogeneous) and a point-mass companion. Within this setting, we have tested our numerical approach against an analytical calculation for this simple case. The semimajor axis' tidal evolution rate was calculated by two methods. One was a direct simulation based on simulating the first body with a mass-spring network. Another, analytical, method was based on a preconceived viscoelastic model of tidal friction (the Kelvin-Voigt model). The numerically computed tidal evolution of the semi-major axis and the spin rate were a direct outcome of the damped mass-spring model. We compared the results obtained by the two methods, and concluded that a mass-spring network can serve as a faithful model of a tidally deformed viscoelastic celestial body. Specifically, we computed the quality function (the ratio of Love number $k_2$ and the quality factor $Q$) from the numerically simulated tidal drift of the semimajor axis. The quality function showed a strong frequency-dependence that is close to the dependence derived analytically for a Kelvin-Voigt sphere by a method suggested by \citet{efroimsky15}. 

While direct estimates for the global shear and bulk rigidity can be derived from the spring constants and spring lengths, the shear viscosity has not been estimated in the literature. Consequently, we were uncertain of our estimates for the shear viscosity and the associated relaxation time (equation \ref{eqn:taurelax}). However, a comparison between our computed quality function (specifically the peak frequency) and that predicted analytically for a Kelvin-Voigt solid suggests that our estimates for the numerically simulated shear viscosity and viscoelastic relaxation time were correct.

 The magnitude of our numerically predicted quality function is about 30\% larger than the one predicted analytically. By comparing simulations performed for two different initial values of the semi-major axis, we concluded that the cause is not the neglect of higher order terms in the quadrupole expansion for the potential. We find the ratio of numerically measured to predicted
 quality function is only weakly dependent on numbers of mass nodes and springs per node simulated but does slowly decrease
 with an increase in the numbers of particles simulated. We have not yet identified the cause of the discrepancy. 
 We suspect that the non-uniformity of the spring network at the body surface could have caused a larger
 than expected  tidal response.  
 Alternatively the simulated shear modulus could be frequency
 dependent and overestimated by its computed static value.

Our study demonstrates that we can directly simulate the tidal evolution of viscoelastic bodies.  
We currently achieve an accuracy of 30\% but with ongoing effort we may improve upon this.
It would be difficult to model this process over  long (Myr or Byr) timescales, because the time step is determined by the number of particles within the body and by the spring constants for the links connecting the particles. This requires the time step to be much shorter than the orbital time scale. Nonetheless, mass-spring models can be used to explore the tidal evolution of inhomogeneous and anisotropic bodies, and to study how their quality functions depend on the rheological properties and internal structure of the body. This fully numerical approach also permits study of tidal phenomena that are not easy to predict analytically~---~such as capture into (and crossing of) spin-orbit and spin-spin resonances, tidally induced orbital evolution, the distribution of tidal heating, 
and the effects of non-linear rheological behaviour.

 \vskip 1.0truein
 \section*{Acknowledgements}
 We thank Beno\^it Noyelles for helpful comments that helped improve the quality of the paper. We also wish to thank Harry Braviner and Moumita Das for helpful discussions, as well as Maciej Kot, Piotr Szymczak, Hanno Rein and Darin Ragozzine. This work was in part supported by the NASA grant NNX13AI27G.

 \appendix

\section{Several basic facts on bodily tides}

  Tidal interactions play a key role in evolution of planetary systems and multiple stars. Their signature is observed, e.g., in the synchronised spin of the Moon, the $\,${\it{pas de deux}}$\,$ of Pluto and Charon, and in the 3:2 spin-orbit resonance of Mercury. Slowly but steadily, tides work to circularise the orbits of planets and moons~---~or, in some cases, to make orbits eccentric.$\,$\footnote{~This happens when the spin of a (tidally-despun) host star is faster than the orbital period of a close planet orbiting it.} Tidal dissipation warms up close-in moons (like the volcanic Io), close-in planets (like bloated Jupiters), and short-period binary stars (which can experience tidal coalescence).

 Referring the reader to \citet{efroimsky15}, \citet{efroimsky13} and references therein for a more detailed introduction, here we provide a minimal kit of ideas and formulae needed to talk about bodily tides.

\subsection{Static tides}

 Consider an extended spherical body of mass $\,M\,$ and radius $\,R\,$, tidally distorted by a perturber of mass $\,M^*\,$ located in an exterior point $\,{\erbold}\,$, so $\,|\erbold|\geq R\,$. In a surface point $\,\Rbold\,$ of the body, the potential due to the perturber is~\footnote{~The reason why summation in the equation (\ref{L1}) goes over $\,l\geq2\,$ is explained, e.g., in \citet[][Eqns. 5 - 11]{efroimsky09}.}
 \ba
 W(\eRbold,\,\erbold)~=~\sum_{{\it{l}}=2}^{\infty}W_{\it{l}}(\eRbold,~\erbold)
 \quad.\quad
 \label{L1}
 \ea
 where the inputs $\,W_{\it{l}}(\eRbold,~\erbold)\,$ are proportional to the appropriate Legendre polynomials $\,P_{\it{l}}(\cos \gamma)\,$, with $\,\gamma\,$ being the angle between the vectors $\,{\erbold}\,$ and $\,\Rbold\,$ pointing from the body centre. The integers $\,l\,$ are termed the {\it{degrees}}.

 The $\,{\emph{l}}$-degree term $\,W_{\it{l}}(\eRbold,\,\erbold)\,$ of the perturber's potential causes a tidal deformation of the perturbed body, assumed to be linear. Then the resulting $\,{\emph{l}}^{~th}$ addition $\,U_{\it{l}}\,$ to the perturbed body's potential is also linear in $\,W_{{l}}\,$:
 \ba
 U(\erbold\,')~=~\sum_{l=2}^{\infty}~U_{l}(\erbold\,')~=~\sum_{l=2}^{\infty}~k_{l}\;\left(\,\frac{R}{r\,'}\,\right)^{l+1}\;W_{l}(\eRbold\,,\;\erbold)\quad,\quad
 \nonumber
 \ea
 $\erbold\,'$ being an exterior point, and $\,k_l\,$ being the static Love numbers.

 Distorting the extended body, the perturber experiences its response in the form of the incremental potential $\,U\,$ taken at the point $\,\erbold\,'\,=\,\erbold\;$:
 \ba
  U(\erbold)~=~\sum_{l=2}^{\infty}~U_{l}(\erbold)~=~\sum_{l=2}^{\infty}~k_{l}\;\left(\,\frac{R}{r}\,\right)^{l+1}\;W_{l}(\eRbold\,,\;\erbold)\quad.\qquad
 \label{L2}
 \ea
 As the perturber is exterior ($\,|\erbold|>R\,$), the quadrupole part of the expansion for the perturbing potential $\,W\,$ is dominant. The same pertains to $\,U\,$.

\subsection{Evolving tides}

 The case of evolving tides is more complicated. Owing to the internal friction, the tidal deformation (and the resulting additional potential $\,U\,$) always lags in time \footnote{~The caveat
 `$\,${\it{in time}}$\,$' is important. Lagging in time does not necessarily imply geometric lagging of the bulge. The lunar orbit being above synchronous, the main (semidiurnal) tide created by
 the Moon on the Earth always leads, not lags. This, however, gets along well with causality.} behind the perturbation $\,W\,$. To take into account different lagging at different frequencies, it
 is necessary to expand both the perturbing potential $\,W\,$ and the response $\,U\,$ in Fourier series. The linearity of response implies that the same frequencies should emerge in both spectra,
 when $\,W\,$ and $\,U\,$ are observed at the same point of space. From the cornerstone work by \citet{kaula64}, it is easy to derive that the Fourier tidal modes read as
 \ba
 \nonumber
 \omega_{\textstyle{_{lmpq}}}&=&(l-2p)\;\dot{\omega}\,+\,(l-2p+q)\;n\,+\,m\;(\dot{\Omega}\,-\,\dot{\theta})\\
  &\approx&(l-2p+q)\;n\,-\,m\;\dot{\theta}
 \label{omega}
 \label{A1}
 \label{L3}
 \ea
 where $\,\theta\,$ and $\,{\bf{\dot{\theta\,}}}\,$ are the rotation angle and rotation rate of the extended body, introduced in the equatorial plane. In neglect of the equinoctial precession,
 $\theta$ can be identified with the sidereal angle. The notations $\,\omega\,$ and $\,\Omega\,$ stand for the perturber's argument of the pericentre and the longitude of the node, as
 seen from the extended body. The formula also includes the mean anomaly $\,{\cal{M}}\,$ and the $\,${\it{anomalistic}}~\footnote{~With $\,a\,$ being the semimajor axis and $\,G\,$ the gravity constant, the mean anomaly $\,{\cal{M}}(t)={\cal{M}}_0(t)+\int^t_{}\,dt\,\sqrt{G(M + M^*)/a^3\,}\,$ renders the anomalistic mean motion as $\,n\equiv {\bf{\dot{\cal{M}}}}={\bf{\dot{\cal{M}}}}_0+\sqrt{G(M + M^*)/a^3\,}\,$. In neglect of external perturbations, $\,{\bf{\dot{\cal{M}}}}_0\approx 0\,$ and the anomalistic mean motion can be approximated with the Keplerian mean motion: $\,n \approx \sqrt{G(M + M^*)/a^3\,}\,$.
 }
 mean motion $\,n\equiv{\bf{\dot{\cal{M}}}}\,$ (with $\,{\cal{M}}=\,0\,$ at the pericentre). Derivation of the expression (\ref{L3}) is explained in Section 4.3 of \citet{efroimsky13}.

 The modes $\,\omega_{\textstyle{_{lmpq}}}\,$ can be of either sign, while their absolute values
 \ba
 \chi_{\textstyle{_{lmpq}}}\,=\,|\,\omega_{\textstyle{_{lmpq}}}\,|~\approx~|\,(l-2p+q)\;n\,-\,m\;\dot{\theta}\,|\,~,
 \label{chi}
 \label{A2}
 \label{L4}
 \ea
 have the meaning of positive definite forcing frequencies of stresses and strains in the distorted body. The Fourier modes are parameterised with the four integers $\,l,\,m,\,p,\,q\,$. The integers $\,l\,$ and $\,m\,$ are the degree and order of the spherical harmonics employed in the
 expansion.$\,$\footnote{~Sometimes $m$ is also referred to as the azimuthal wavenumber (\citealt{ogilvie14}).}

 The dynamical analogue to the formula (\ref{L1}) is:
 \ba
 W(\eRbold,\,\erbold,\,t)=\sum_{l=2}^{\infty} W_{l}(\eRbold,\,\erbold,\,t)
  =\sum_{lmpq}~W_{{lmpq}}(\eRbold,\,\erbold,\,t)~~~,
  \label{L5}
 \ea
 where a term $\,W_{{lmpq}}\,$ is proportional to $\,\cos\left(\omega_{lmpq}\,t\,+\,.\,.\,.\,\right)\,$, with ellipsis denoting some phase:
 \ba
 W_{{lmpq}}(\eRbold,\,\erbold,\,t)=A_{{lmpq}}(\eRbold,\,\erbold,\,t)\;\cos\left(\omega_{lmpq}\,t\,+\,.\,.\,.\,\right)\quad.\quad
 \label{L6}
 \ea
 Both the static formula (\ref{L1}) and its dynamical analogue (\ref{L5}) render the value of the perturbing potential at a surface point $\,\Rbold\,$.

 Writing down a dynamical analogue to the static expression (\ref{L2}) turns out to be a highly nontrivial problem. Above we stated that, owing to the linearity of the problem, the spectrum of
 $\,U\,$ should contain the same frequencies as that of $\,W\,$, {\it{provided both $\,U\,$ and $\,W\,$ are observed at the same point of space}}. Therefore, a Fourier series for $\,U\,$ would
 contain terms proportional to $\,\cos\left(\omega_{lmpq}\,t\,+\,.\,.\,.\,\right)\,$, had it been written for the (evolving in time) value of  $\,U\,$ at the same surface point $\,\Rbold\,$. We
 however are interested in the values of $\,U\,$ in a different point, the point $\,\erbold\,$ where the moving perturber is located.  There, the spectrum of  $\,U(\erbold,\,t)\,$ will be richer than that of $\,W(\eRbold,\,\erbold,\,t)\;$, and will be parameterised with six indices $\,lmpqhj\;$:
 \ba
 U(\erbold,\,t)~=~\sum_{{\it{l}}=2}^{\infty}U_{{l}}(\erbold)~=~\sum_{lmpqhj}~U_{{lmpqhj}}(\erbold,\,t)
 \quad,\quad
 \label{L7}
 \ea
 see \citet[][Sections 7 \& 8]{efroimsky12a}. As was pointed out by \citet{kaula64}, $\,U(\erbold,\,t)\,$ contains a secular part~---~and that part is parameterised with the four indices
 $\,lmpq\;$:
 \ba
 \langle \,U(\erbold,\,t)\,\rangle~=~\sum_{{\it{l}}=2}^{\infty}\langle\,U_{l}(\erbold,\,t)\,\rangle~=~\sum_{lmpq}~\langle U_{{lmpq}}(\erbold) \rangle
 \quad,\quad
 \label{L8}
 \ea
 where the angular brackets $~\langle\,.\,\,.\,.\,\rangle~$ denote time-averaging, and the terms on the right-hand side are given by
 \begin{gather}
 \nonumber
 \langle U_{{lmpq}}(\erbold)\rangle~= \qquad\qquad\qquad\qquad\qquad\qquad\qquad\qquad\qquad\qquad\\
 k_l(\omega_{lmpq})~\cos\epsilon_l(\omega_{lmpq})~\left(\,\frac{R}{r}\,\right)^{l+1}\;A_{lmpq}(\eRbold\,,\;\erbold)\quad,\quad
 \label{L9}
 \end{gather}
 where $\,A_{lmpq}\,$ are the magnitudes from the formula (\ref{L6}), while $\,k_l(\omega_{lmpq})\,$ and $\,\epsilon_l(\omega_{lmpq})\,$ are the degree-{\emph{l}} dynamical Love numbers and
 phase lags written as functions of the Fourier modes.

\subsection{The secular part of the tidal torque acting on the spin of the extended body}

 The negative gradient of the secular potential (\ref{L8}) renders the secular part of the orbital torque wherewith the extended body is acting on the perturber. An equal but opposite torque is
 acting on the extended body and is influencing its spin. The polar component of the secular torque reads as
 \ba
 \langle\,{\cal{T}}^{(z)}\,\rangle~=~\sum_{{\it{l}}=2}^{\infty}\langle\,{\cal{T}}_l^{(z)}\,\rangle~=~\sum_{lmpq}~\langle\,{\cal{T}}_{lmpq}^{(z)}\,\rangle
 \quad,\quad
 \label{L10}
 \ea
 where
 \ba
 \nonumber
 \langle\,{\cal{T}}_{lmpq}^{(z)}\,\rangle~=~\qquad\qquad\qquad\qquad\qquad\qquad\qquad\qquad\qquad\qquad\\
 k_l(\omega_{lmpq})~\sin\epsilon_l(\omega_{lmpq})~\left(\,\frac{R}{r}\,\right)^{l+1}m\;A_{lmpq}(\eRbold\,,\;\erbold)\quad.\quad
 \label{L11}
 \ea
 We see that an $\,lmpq\,$ component of the torque may be either decelerating or accelerating the spin, dependent upon the sign of the phase lag $\,\epsilon_l(\omega_{lmpq})\,$~---~which always coincides with the sign of the Fourier mode $\,\omega_{lmpq}\,$.

\subsection{The quality function (``kvalitet")} \label{sec:kva}

 The product $~k_l(\omega_{lmpq})\;\sin\epsilon_l(\omega_{lmpq})~$ is sometimes termed as $\,${\it{the quality function}}$\,$ \citep{makarovMoon,efroimsky15} or
 $\,${\it{kvalitet}}$\,$ \citep{makarovSemiliquid,frouard}. In the literature, it is conventional to write it as
 \ba
 k_l(\omega_{lmpq})\;\sin\epsilon_l(\omega_{lmpq})~=~\frac{k_l(\omega_{lmpq})}{Q_l(\omega_{lmpq})}~\,\mbox{Sgn}\,\omega_{lmpq}\quad,
 \label{L12}
 \ea
 where the quality factors are introduced via
 \ba
 \frac{1}{Q_l(\omega_{lmpq})}\,=\,|\,\sin\epsilon_l(\omega_{lmpq})\,|\quad,
 \label{L13}
 \ea
 and where it is taken into account that the sign of a phase lag $\,\epsilon_l(\omega_{lmpq})\,$ always coincides with the sign of the Fourier mode $\,\omega_{lmpq}\;$ \citep[e.g.,][]{efroimsky13}.

\subsection{Which terms are leading, and when\label{1.5}}

 As the perturber is exterior ($\,|\erbold|>R\,$), the quadrupole part of the expansion for the perturbing potential $\,W\,$ is dominant. The quadrupole part comprises all the terms with
 $\,l=2\,$. For low inclination and eccentricity, the largest terms in the expansions (\ref{L5}), (\ref{L7}), and (\ref{L10}) are those with  $\,\{lmpq\}=\{2200\}\,$. They correspond to the
 so-called semidiurnal Fourier mode
 \ba
 \omega \, \, \equiv ~\omega_{2200}\,=~2~(n~-~\dot\theta) \quad.\quad
 \label{eqn:omega}
 \label{L14}
 \ea
 When the semidiurnal, or any other $\,lmpq\,$ term is leading in the expansion for $\,W\,$, the corresponding $lmpq$ term is leading also in the expansion for the additional tidal potential $\,U\,$.
 Up to some reservation, this is true also for the expansions of the tidal torque. A reservation comes from the fact that an $\,lmpq\,$ term in the expansion for the torque contains as a
 multiplier the sine of the phase lag $\,\epsilon_l(\omega_{lmpq})\,$. For example, in the case of small inclination $\,i\,$ and eccentricity $\,e\,$, the semidiurnal part of the polar torque
 operating on the spin of the perturbed body reads as \citep{efroimsky12a}:
 \begin{gather}
 \label{L15}
 {\cal{T}}^{(z)}_{2200}\,= \quad
 ~\\
 \nonumber
 \frac{3}{2} G {M^*}^2 \frac{R^5}{a^6} k_2(\omega_{2200})\;\sin\epsilon_2(\omega_{2200})\,+\,O(e^2\,\epsilon)\,+\,O(i^2\,\epsilon)~~.
 \end{gather}
 The quality function $\,k_l(\omega_{lmpq})\;\sin\epsilon_l(\omega_{lmpq})\,$ continuously goes through zero (and changes its sign) when the $\,lmpq\,$ spin-orbit resonance is transcended, i.e.,
 when $\,\omega_{lmpq}\,$ goes through zero. So, when a rotator is trapped into an $\,lmpq\,$ spin-orbit resonance, the quality function stays zero; so the Fourier mode $\,\omega_{lmpq}\,$
 contributes nothing to the torque. Specifically, in the case of synchronous rotation (known as $\,${\it{the 1:1 spin-orbit resonance}}), the mode $\,\omega_{2200}\,$ vanishes~---~and so does the
 semidiurnal term of the torque. In the resonance, therefore, it is the higher-than-semidiurnal terms that are leading.

 This ``acceding of leadership" in resonances, along with its physical consequences for binaries, is described in detail in \citet{makarov&efroimsky} and \citet{makarovetal}. Here we shall only mention two simple examples.
 Since the Moon is synchronised, the semidiurnal input into the torque acting on its spin is zero. It is then the other components (mainly, the term with $\,\{lmpq\}\,=\,\{2201\}\,$) that define
 the tidal response of the Moon and influence its libration in longitude \citep{frouard}. As another example, take Mercury in its 3:2 resonance \citep{noyelles14}. For this planet, the
 $\,\{lmpq\}\,=\,\{2201\}\,$ input into its tidal response is zero, and it is the semidiurnal mode that overwhelmingly defines the tidal response and plays a crucial role in longitudinal libration
 \citep{makarov}.

\subsection{Tidally generated secular orbital evolution} \label{sec:sec}

 The tidal potential (Eq.\ref{L8}) should be inserted into the Lagrange- or Delaunay- type planetary equations, to calculate the secular evolution of the orbit.$\,$\footnote{~Generally, the two bodies should be treated on equal footing, so this potential should be amended with a similar potential wherewith the extended body is acted upon due to the tides it is exerting on the perturber.}$\,$ It then turns out after some algebra that the secular orbital evolution is determined mainly by the quality function $~k_l(\omega_{lmpq})\;\sin\epsilon_l(\omega_{lmpq})~$, with the cosine of the lag playing a very marginal role.$\,$\footnote{~The evolution of the argument of the pericentre $\,\omega\,$, the longitude of the node $\,\Omega\,$, and the mean motion $\,{\cal{M}}\,$ depends overwhelmingly on
 $~k_l(\omega_{lmpq})\;\sin\epsilon_l(\omega_{lmpq})~$, and also contains terms with
 $~k_l(\omega_{lmpq})\;\cos\epsilon_l(\omega_{lmpq})~$. The latter terms, though, are very small. It can also be shown that the secular drift of the semimajor axis $\,a\,$, eccentricity $\,e\,$, and inclination $\,i\,$ is defined exclusively by the quality functions $~k_l(\omega_{lmpq})\;\sin\epsilon_l(\omega_{lmpq})~$, with no terms containing
 $~k_l(\omega_{lmpq})\;\cos\epsilon_l(\omega_{lmpq})~$.}

 However, in some situations, approximate secular evolution can be calculated via the tidal torque. For example, consider an orbit with no inclination relative to the equator of the extended body. Using the expression
 \ba
  \;L^{(orb)}\;=\;\frac{M M^*}{M + M^*} \sqrt{G (M + M^*)} \sqrt{a(1-e^2)}
 \label{L16}
 \ea
 for the orbital angular momentum, and setting there $\,e=0\,$, we can use the conservation of the angular momentum $~\dot{L}^{(orb)}\,=\,-{\cal{T}}^{(z)}~$ to derive the evolution rate of the semimajor axis:
 \begin{eqnarray}
  \frac{\dot a}{n\,a} &=&~-\frac{2\,{\cal{T}}^{(z)} \,a}{G M^* M}
  \nonumber\\
  \label{L17}\\
  \nonumber
 &=& -\;3 \left( \frac{M^*}{M} \right)\left( \frac{R}{a} \right)^5\;k_l(\omega_{lmpq})\;\sin\epsilon_l(\omega_{lmpq}) \quad.\quad
 \end{eqnarray}
 The expression (\ref{L17}) is a semidiurnal approximation. As was mentioned in Subsection \ref{1.5}, this approximation is valid everywhere except in the 1:1 resonance where the semidiurnal term vanishes.

{}

\end{document}